\documentstyle[11pt]{article}

\begin{document}

\bibliographystyle{prsty}


\title{Simulated Dynamics of Underpotential Deposition of Cu with
Sulfate on Au(111)}

\author{Gregory Brown$^{\rm a,b}$,Per Arne Rikvold$^{\rm a,b}$,
        M. A. Novotny$^{\rm b}$,\\ and Andrzej Wieckowski$^{\rm c}$\\
{~}\\
$^{\rm a}$Center for Materials Research and Technology\\
and Department of Physics\\ 
Florida State University, Tallahassee, FL 32306-4052\\
$^{\rm b}$~Supercomputer Computations Research Institute\\ 
Florida State University, Tallahassee, FL 32306-4130\\
$^{\rm c}$~Department of Chemistry\\ 
and Frederick Seitz Materials Research Laboratory\\ 
University of Illinois, Urbana, IL 61801\\
}
\maketitle

\begin{abstract}
Numerical studies of lattice-gas models are well suited to describe
multi-adsorbate systems. One example is the underpotential deposition
of Cu on Au(111) in the presence of sulfuric acid.  Preliminary
results from dynamic Monte Carlo simulations of the evolution of the
adsorbed layer during potential-step experiments across phase
transitions are presented for this particular system. The simulated
current profiles reproduce a strong asymmetry seen in recent
experiments. Examination of the microscopic structures that occur
during the simulated evolution processes raises questions that need to
be investigated by further experimental and theoretical study.

{~}
Multi-adsorbate systems.
Lattice-gas model.
Butler-Volmer approximation.
Dynamic Monte Carlo simulation.

\end{abstract}

\newpage

\section{Introduction}

Underpotential deposition (UPD) provides a window of electrode
potentials in which a monolayer or less of one metal can be deposited
onto the surface of another metal.  Recently, experiments that probe
the dynamical properties of the adsorption process \cite{HOLZ94} have
also added to our knowledge of UPD monolayers. Here we present
preliminary results for dynamical simulations of a microscopic model
which reproduce the results of potential-step experiments and make
microscopic predictions that may be testable by modern {\em in-situ}
experimental techniques, such as scanning-probe microscopy and
synchrotron X-ray scattering.

The UPD of copper onto gold single-crystal electrodes in the presence
of sulfate is a particularly well studied model system. See, for
example, the review of the experimental literature in
Ref.~\cite{ZHAN96}.  For this system cyclic voltammograms, like the
one shown in Fig.~1, show two current peaks corresponding to two
different transitions in the surface structure
\cite{SCHU76}. Scanning-tunneling microscopy (STM)
\cite{HACH91,MAGN91} and low-energy electron diffraction (LEED)
\cite{ZHAN96} studies indicate that these transitions occur between a
full monolayer (ML) of Cu at more negative potentials, an ordered
$(\sqrt{3} \times \sqrt{3})$ mixed copper and sulfate phase at
intermediate potentials, and a disordered low-coverage phase at more
positive potentials.  {\em In situ\/} X-ray scattering confirms that
the ordered phase consists of a 2/3 ML of copper in a honeycomb
pattern, with a 1/3 ML of sulfate occupying the remaining adsorption
sites \cite{TONE95}, as originally proposed by Huckaby and Blum
\cite{HUCK91}.

An effective theoretical approach for understanding a multi-species
layer chem\-i\-sorb\-ed on a surface is to study a statistical-mechanical
lattice-gas model numerically. Such models describe the adsorbate
layer by a Hamiltonian which gives the energies of different adsorbate
configurations. Early studies \cite{HUCK91,RIKV88,BLUM90} together
with later applications to problems such as the adsorption of urea on
Pt(100) \cite{GAMB93,RIKV95} and the UPD of Cu with sulfate on Au(111)
\cite{ZHAN96,HUCK91}, in or near equilibrium, are discussed in recent
review articles \cite{RIKV91,RIKV96,BLUM96}.  In particular, that work
provided estimates for effective lateral interaction energies, based on
comparison of the theoretical predictions with experimentally observed
equilibrium adsorption isotherms and voltammetric currents for small
potential sweep rates.

Dynamical properties of the copper UPD system have recently been
investigated by H{\"o}lzle, Retter, and Kolb (HRK) \cite{HOLZ94}.
They measured current transients in potential-step experiments
performed at both transitions and with both positive-going and
negative-going steps. In this article we report current transients and
time-dependent microscopic adlayer structures observed in dynamic
Monte Carlo simulations of a model UPD system under conditions
intended to reproduce those experiments.  In a previous study
\cite{RIKV97a}, our model agreed qualitatively with the experiments
only for steps across one transition, the mixed-disordered phase
transition marked ${\rm A}$ and ${\rm A}'$ in Fig.~1. Reproducing the
behavior seen for potential steps across both transitions represents a
significant refinement of the lattice gas model. Our microscopic
predictions indicate the desirability of time-resolved, structurally
sensitive, {\em in situ\/} experimental studies of this system.

\section{Lattice-gas Model}

Our theoretical lattice-gas model treats copper and sulfate as
interacting particles that compete for the same adsorption sites. For
the Au(111) surface this is a triangular lattice of threefold hollow
sites \cite{TONE95}. The model is defined by the effective lattice-gas
Hamiltonian,
\begin{eqnarray} 
\label{eq:lgham}
{\mathcal H} & = & - \sum_n 
\left[ 
 \Phi_{\rm CC}^{(n)} \sum_{\langle ij \rangle}^{(n)} c_i^{\rm C} c_j^{\rm C}
+\Phi_{\rm CS}^{(n)} \sum_{\langle ij \rangle}^{(n)} 
           \left(
              c_i^{\rm C} c_j^{\rm S} 
              + c_i^{\rm S} c_j^{\rm C} 
           \right)  
+\Phi_{\rm SS}^{(n)} \sum_{\langle ij \rangle}^{(n)}  c_i^{\rm S} c_j^{\rm S} 
\right]  \\
& & -\Phi_{\rm SS}^{(t)} \sum_{\triangle} c_i^{\rm S} c_j^{\rm S} c_k^{\rm S}
    -\sum_i \bar{\mu}_{\rm C} c_i^{\rm C}  
    -\sum_i \bar{\mu}_{\rm S} c_i^{\rm S}  \nonumber
\;.
\end{eqnarray}
Here $c_i^{\rm X}$ equals unity when lattice site $i$ is occupied by
species ${\rm X}$, with ${\rm X}={\rm C}$ for copper and ${\rm X}={\rm
S}$ for sulfate; otherwise $c_i^{\rm X}$ equals zero. This third state
represents sites solvated by water. The effective interaction for
$n$-th neighbor adsorbates ${\rm X}$ and ${\rm Y}$ is $\Phi_{\rm
XY}^{(n)}$, and $\sum_{\langle ij \rangle}^{(n)}$ represents the sum
over all $n$-th order neighbor pairs. We also include a three-body
interaction, $\Phi_{\rm SS}^{(t)}$, for equilateral triangles of
sulfate adsorbed on second-neighbor sites. The sum over all possible
triangles is denoted by $\sum_{\triangle}$. Finally, $\bar{\mu}_{\rm
X}$ is the electrochemical potential for ${\rm X}$, and $\sum_i$ is
the sum over all lattice sites. Attractive interactions are
represented by positive quantities, and positive electrochemical
potentials indicate a preference for adsorption in the absence of
lateral interactions.

The quantities $\Phi_{\rm XY}^{(n)}$ in the lattice-gas model are
effective parameters representing the joint effect of competing,
unspecified interaction potentials. Ground-state calculations can be
used to determine multi-dimensional regions of parameter space that
give phase diagrams that agree with the phases observed
experimentally. Fitting the model to a particular system involves an
iterative process of selecting experiments and adjusting the
parameters to achieve improved agreement. The expectation is that the
model will more accurately represent the real chemical system as it
correctly reproduces and predicts an ever wider range of experimental
observations. Detailed discussion of the methods used to estimate the
effective interaction constants are given in
Refs.~\cite{ZHAN96,RIKV95,RIKV91}.

The effective interactions used in the present study are illustrated
in Fig.~2. There, and elsewhere in this paper, unoccupied adsorption
sites are represented by an open circle ({\Large$\circ$}), sites
occupied by copper are represented by a filled circle
({\Large$\bullet$}), and those occupied by sulfate are represented by
a triangle ({$\triangle$}). These interactions have been adapted from
those of Ref.~\cite{ZHAN96}. Changes in the sulfate-sulfate
interactions removed from the ground-state diagram three
experimentally unobserved low-temperature phases which were overlooked
in Ref.~\cite{ZHAN96}. Strengthening the attractive next-nearest
neighbor copper-copper interaction, $\Phi_{\rm CC}^{(2)}$, was
required to make the phase transition between the copper monolayer and
the mixed phase (peaks ${\rm B}$ and ${\rm B}'$ in Fig.~1) first order
and produce current transients for this transition in qualitative
agreement with experiments (described in Sec.~4).

The ground-state diagram for the interactions of Fig.~2 is presented
in Fig.~3, and the geometries of the main observed phases are
illustrated in Fig.~4.  Here the phases are labeled by the length of
their basis vectors and subscripted (superscripted) by the fraction of
sites filled by copper (sulfate).

The electrochemical potential of species ${\rm X}$ is controlled
through its bulk solution activity and the electrode potential,
$E$. In the weak-solution limit, the activity equals the solute
molar concentration $[{\rm X}]$. Then the electrochemical potential,
$\bar{\mu}_{\rm X}$, is
\begin{equation}
\label{eq:echempot}
\bar{\mu}_{\rm X} = 
   \mu_{\rm X}^0 + RT\ln{\frac{[{\rm X}]}{[{\rm X}]^0}}
                 - z_{\rm X}FE, 
\end{equation}
where $z_{\rm X}$ is the effective electrovalence of ${\rm X}$, $F$ is
Faraday's constant, $R$ is the molar gas constant, $T$ is temperature,
and the reference state is ${\mu}_{\rm X}^0$, $[{\rm X}]^0$, and
$E^0=0$.  For fixed $[{\rm S}]$ and $[{\rm C}]$,
Eq.~(\ref{eq:echempot}) for $\bar{\mu}_{\rm S}$ and $\bar{\mu}_{\rm
C}$ form the parametric representation of a line in the ground-state
diagram. The electrode potential $E$ is the parameter that determines
the location along this {\em isotherm\/}, whose slope is $z_{\rm
S}/z_{\rm C}$. We have chosen an isotherm such that the potential
difference between the two transitions is the same as observed by
HRK. Its location is given by the dotted line
in Fig.~3 for $0 \le E \le 350 {\rm mV}$ vs Ag/AgCl.

\section{Dynamic Monte Carlo Method}

The equilibrium properties of a lattice-gas Hamiltonian can often only
be practically evaluated using Monte Carlo methods, by which
thermodynamic average values are calculated from a representative
sampling of the equilibrium states. This sampling is generated using
the principle of detailed balance: in equilibrium the rate of going
from configuration $A$ to configuration $B$ is equal to the rate of
going from $B$ to $A$. Many Monte Carlo techniques generate new
configurations through a series of Monte Carlo steps, each step
consisting of a proposed small change in the configuration which is
accepted or rejected according to an acceptance probability that must
satisfy detailed balance.

A lattice-gas Hamiltonian merely provides the free energies
corresponding to different geometric configurations of the adsorbate
layer. In contrast to a quantum-mechanical Hamiltonian, it does
not possess an intrinsic dynamic. However, when Monte Carlo steps are
associated with a discrete unit of time, the simulation can be viewed
as a ``movie'' of a dynamic process.

Many different Monte Carlo algorithms can be designed to describe the
equilibrium properties of a particular system. In general such an
algorithm does not realistically describe the dynamics of a particular
chemical system. But if care is taken to include only dynamically
realistic moves in the algorithm, it is possible to construct a Monte
Carlo dynamic which simulates that of the real system under
consideration. This should be the case when the real dynamics is
well approximated by a stochastic process on the timescales of
interest.  The example relevant to our present discussion is the
thermally activated motion of particles adsorbing onto, desorbing
from, and diffusing between adsorption sites on the electrode
surface. This view of Monte Carlo simulation can be combined with the
lattice-gas Hamiltonian to construct a stochastic dynamic describing
the individual motions of specific adsorbate particles. An
introduction to Monte Carlo methods for equilibrium and nonequilibrium
problems in surface electrochemistry is given in Ref.~\cite{CHAP}.

Here we construct a microscopic dynamic which includes adsorption,
desorption, and one-step lateral diffusion of both adsorbate
species. The rate of each process is determined using a simple
Arrhenius picture, with all processes having the same pre-exponential
factor $\nu_0$.  The free energy associated with an adsorbed particle
of species ${\rm X}$ occupying site $i$ for a specific configuration
of neighbors is given by
\begin{equation}
{\mathcal H}_i({\rm X};\Gamma) = - \sum_n 
\left[ 
 \Phi_{\rm XC}^{(n)} \sum_{j(i)}^{(n)} c_j^{\rm C}
+\Phi_{\rm XS}^{(n)} \sum_{j(i)}^{(n)} c_j^{\rm S}
\right]
-\delta_{\rm X,S}\Phi_{\rm SS}^{(t)}\sum_{\triangle(i)} c_j^{\rm S} c_k^{\rm S}
-\bar{\mu}_{\rm X }
\;,
\end{equation}
where $\sum_{j(i)}^{(n)}$ represents the sum over all adsorption sites
$j$ that are $n$-th neighbors of site $i$, and $\sum_{\triangle(i)}$
is the sum over all second-neighbor equilateral triangles involving
site $i$. The factor $\delta_{\rm X,S}$ is unity when ${\rm X}={\rm
S}$ and zero otherwise. The index $\Gamma$ runs over all possible
arrangements of neighboring adsorbate particles within the maximum
interaction range from site $i$. When the site is empty and the ion is
in solution, the free energy is defined to be zero. Using ${\rm X}=0$
to denote an unoccupied site, ${\mathcal H}_i(0;\Gamma)=0$, regardless
of the arrangement of the neighbors.

The energy barriers are assumed to vary according to the Butler-Volmer
approximation \cite{BARD}. Specifically, consider the
adsorption-desorption processes for a particular site. A schematic
free-energy surface is shown in Fig.~5(a), assuming three different
arrangements of neighbors, $\Gamma$. If ${\mathcal H}_i({\rm
X};\Gamma)=0$, there is no preference toward adsorption, but there is
still a transition-state free energy, $\Delta^*_0({\rm X})$.  The
transition-state free energy for this case is used as the reference
value for the adsorption transition energy for all configurations of
the neighbors. If a site has no neighbors within the maximum
interaction range, the free energy of the adsorbed particle is
controlled by the electrochemical potential and equals
$-\bar{\mu}_{\rm X}$.  In the Butler-Volmer approximation, the
transition energy for adsorption to that site is then lowered from
$\Delta^*_0({\rm X})$ by an amount $\alpha\bar{\mu}_{\rm X}$. Here
$\alpha$ is the symmetry factor, which is often close to $1/2$.  For
the present model, lateral interactions with neighboring ions are also
important. If the net lateral interactions are attractive, the free
energy corresponding to the adsorbed state lies below $-\bar{\mu}_{\rm
X}$, as illustrated by the lowest adsorbed free-energy level in
Fig.~5(a). If the net lateral interactions are repulsive, the free
energy is raised above $-\bar{\mu}_{\rm X}$. Our generalization of the
Butler-Volmer approximation uses the total free energy gained to
calculate the free-energy barrier for adsorption, specifically
\begin{equation}
\Delta^*_i({\rm X};\Gamma) = \Delta^*_0({\rm X}) 
                           + \alpha {\mathcal H}_i({\rm X};\Gamma) \;.
\end{equation}
Using the same transition-state energy $\Delta^*_i({\rm X};\Gamma)$
for the adsorption and desorption process is essential for Monte Carlo
simulations in order to satisfy detailed balance and thus ensure
eventual approach to thermodynamic equilibrium. In the model developed
here, an ion approaching the surface from the solution sees different
free-energy barriers for sites with different arrangements of
neighbors. Adsorption on a site is faster when its neighbors help
lower the free energy of the adsorbed state; conversely, desorption
from such a site is relatively slow. For multi-component adsorption,
the adsorption of one species may be much faster than that of the
other at a particular site.

For lateral diffusion from one adsorption site to another, a slightly
modified approach is needed. The situation for two neighboring sites,
both with favorable adsorption free energies, is shown in Fig.~5(b).
The difficultly is that detailed balance requires a single
transition-state free energy, $\tilde\Delta^*_{ij}({\rm X};\Gamma)$,
between the two sites. An obvious solution would be to adjust the
energy barrier by the average of the free energies for the two sites,
$({\mathcal H}_i({\rm X};\Gamma)+{\mathcal H}_j({\rm X};\Gamma))/2$
\cite{VATT96}. Unfortunately this could lead to ${\mathcal H}_i({\rm
X};\Gamma) < \tilde\Delta^*_{ij}({\rm X};\Gamma) < {\mathcal H}_j({\rm
X};\Gamma)$, which violates the usual definition of a transition-state
free energy. To avoid this, we use the least favorable adsorption free
energy for calculating the transition-state free energy for diffusion
between nearest-neighbor sites $i$ and $j$:
\begin{equation}
\tilde\Delta^*_{ij}({\rm X};\Gamma) = \tilde\Delta^*_0({\rm X}) 
+ \frac{1}{2} {\rm Max}({\mathcal H}_i({\rm X};\Gamma),{\mathcal H}_j({\rm X};\Gamma)) \;.
\end{equation}
Violations of the transition-state picture are minimized because they
involve very unfavorable sites, which are only rarely occupied.  The
barrier for diffusion to each neighbor is calculated individually;
diffusion into a less favorable site will be relatively slow.
Diffusion to an occupied site is forbidden and has a rate identically
zero. For our simulations, we choose the transition-state free
energies as $\Delta^*_0({\rm C})=50\,{\rm kJ/mol}$ for copper
adsorption, $\Delta^*_0({\rm S})=55\,{\rm kJ/mol}$ for sulfate
adsorption, and $\tilde\Delta^*_0({\rm C})=\tilde\Delta^*_0({\rm
S})=35\,{\rm kJ/mol}$ for diffusion of both species. These choices
make adsorption faster for copper than sulfate and allow relatively
long diffusion paths across the surface since diffusion is faster than
adsorption. We take the symmetry factor $\alpha=1/2$.

The rates described above can be used to implement a rejection-free
Monte Carlo algorithm \cite{CHAP,NOVO95}. A list of all the single-ion
processes is maintained. Each process is weighted by its rate, so that
making a random choice from the list is biased towards the faster
processes.  Each simulation step consists of choosing a process from
the weighted list, making the appropriate change to the system, and
recalculating the weights for the list. At each step the simulation
clock is updated by a random number that represents the elapsed time
since the last move. For a given configuration the overall rate for
the system to change by some microscopic process is the sum of all the
individual microscopic process rates. Since the decay of a particular
state has a constant rate, the elapsed time for the actual change has
an exponential distribution.  Results for a simplified model were
presented in Ref.~\cite{RIKV97}.

Our simulations were conducted on $30 \times 30$ parallelograms of a
triangular lattice with periodic boundary conditions. A particular
potential-step simulation involves equilibration for a fixed time at
one potential. Then, at time $t=0$, the applied potential is changed
to its final value, and measurements of the partial coverage due to
each species are made at fixed simulation times after the
quench. Results reported here are averaged over at least 1000
individual simulations. Currents are calculated by numerical
differentiation of the coverage with respect to time, with
electrovalences of $z_{\rm S}=-0.9$ and $z_{\rm C}=+1.54$
\cite{ZHAN96}. The natural simulation time unit is Monte Carlo steps
per site (MCSS), but choosing $\nu_0=2\times10^9\;{\rm MCSS}\,{\rm
s}^{-1}$ makes the scale of simulated currents agree roughly with
experiments. The rejection-free algorithm is essential, since
achieving a time scale of $10^9\;{\rm MCSS}$ would be
prohibitive for a traditional Monte Carlo algorithm. Even with the
rejection-free algorithm, some of the individual experiments took $15$
minutes to simulate on a DEC-Alpha workstation.

\section{Results}

The agreement between our simulations and the experiments of HRK
\cite{HOLZ94} is encouraging (see Figs.~6 and 7).  The current
profiles, consisting of a rapid initial decay followed by a broad
maximum at a later time, are characteristic of the decay of an
adsorbed layer that is metastable after a potential step across a
first-order phase transition. The initial current transient
corresponds to the rapid relaxation to a long-lived metastable state
which has only a small concentration difference from the adsorbate
layer as it existed before the potential was changed. The second,
broader current maximum occurs as domains of the equilibrium phase
nucleate and grow. This behavior is seen for three of the four
potential steps.

The simulated current transients for potential steps around the
transition between the $(1\times 1)^0_1$ and
$(\sqrt{3}\times\sqrt{3})^{1/3}_{2/3}$ phases, labelled ${\rm B}$ and ${\rm
B}'$ in Figs.~1 and 3, are shown in Fig.~6.  Both reflect one-step
nucleation and growth mechanisms.  Following the trend seen in the
experimental data, the broad maximum becomes stronger and
the maximum shifts to earlier times as the size of the step is
increased. However, the time corresponding to the second maximum
appears to depend more strongly on the step size than observed in the
experiments.

The current transients observed for the transition between the
$(\sqrt{3}\times\sqrt{3})^{1/3}_{2/3}$ and the disordered low-coverage
phases, labelled ${\rm A}$ and ${\rm A}'$ in Figs.~1 and 3, are very
different for positive and negative potential steps, as seen in
Fig.~7. This was also observed by HRK. When the potential is increased
from the ordered-phase region into the disordered-phase region, a step
corresponding to ${\rm A}'$, the current profile has the broad
maximum, similar to those shown in Fig.~6. Again, the qualitative
trend seen for the strength and location of the second current maximum
as the size of the step is changed agrees with the experiments. In
contrast to our previous simulations, which were performed for a
system with different interactions \cite{RIKV97a}, this current is
caused by the decay of the metastable
$(\sqrt{3}\times\sqrt{3})_{2/3}^{1/3}$ phase to the
$(\sqrt{3}\times\sqrt{3})_{1/3}^{1/3}$ phase, which is also
metastable. The decay of this second metastable phase to the final
low-coverage phase takes place at much later times, but causes a
current transient with a much smaller amplitude because the copper and
sulfate currents have opposite sign. Both metastable phases decay by a
nucleation and growth mechanism.

The current profiles observed for potential steps in the negative
direction, corresponding to peak ${\rm A}$, are qualitatively
different than the other profiles.  They are monotonically decreasing
and do not correspond to a simple functional form. A sequence of
configurations after a potential step to $20\,{\rm mV}$ below the
transition is shown in Fig.~8.  Because the applied potentials of HRK
have been used, the equilibration occurs in a disordered phase with a
relatively high sulfate coverage of approximately $0.25$ ML. After the
potential step, copper adsorbs quickly onto the surface. The sulfate
already present collapses with this copper to form a domain of the
$(\sqrt{3}\times\sqrt{3})_{1/3}^{1/3}$ phase. Later this fills in with
copper, forming the equilibrium $(\sqrt{3}\times\sqrt{3})_{2/3}^{1/3}$
phase. The adsorption of sulfate is an even slower process, which
controls the shrinking of the domains of unoccupied sites. This
process is different from our previous simulations \cite{RIKV97a},
where no domain formation was observed. Instead, copper adsorbed into
a loose network with open sites favorable for sulfate adsorption.

The fact that different dynamical paths can give current profiles
qualitatively similar to those observed experimentally indicates that
more study of the copper UPD system is desirable. Particularly useful
would be time-resolved experiments that can probe the microscopic
structure of the adsorbed layer. For the simulations, the factors
controlling the different dynamical behaviors are not clearly
understood and a more thorough investigation is needed.

\section{Conclusions}

Preliminary dynamic Monte Carlo simulation results are reported for
potential step experiments for UPD of copper with sulfate on
Au(111). The simulated current transients display the same general
characteristics as those observed experimentally by H{\"o}lzle,
Retter, and Kolb \cite{HOLZ94}. The current profiles for potential
steps across the transition separating the copper monolayer and the
ordered mixed phase (the transition marked ${\rm B}$ and ${\rm B}'$ in
Fig.~3) show a sharp initial transient followed by a broad second
maximum that shifts to earlier times and larger amplitudes as the size
of the step is increased. This contrasts with the asymmetry of the
current profiles for potential steps across the transition separating
the ordered mixed phase from the disordered low-coverage phase (the
transition marked ${\rm A}$ and ${\rm A}'$ in Fig.~3), where the
second maximum is only seen for positive-going steps. Negative
potential steps produce a monotonically decreasing current. We find
that reproducing the profiles of HRK is not sufficient for determining
the underlying dynamical process. For two slightly different models we
have observed different processes occurring after the step. For both
the negative-going and positive-going steps at this transition (${\rm
A}$ and ${\rm A}'$), an intermediate ordered phase with only $1/3$
monolayer of copper is observed in the model presented here that was
not observed for an earlier model. In addition, domain formation is
seen for negative steps. This was not observed in our simulations of
the earlier model. Experiments which can dynamically monitor the
microscopic structure of the adsorbate layer during the relaxation
after the potential step would be useful in guiding further model
development.

\section*{Acknowledgements}
We acknowledge useful discussions with M. Kolesik.
This research was supported by Florida State University through the
Center for Materials Research and Technology and the Supercomputer
Computations Research Institute (DOE Contract No. DE-FC05-85ER-25000),
and by NSF Grant No. DMR-963483. Work at the University of Illinois
was supported by NSF Grant No. CHE-97-00963 and by the Frederick Seitz
Materials Research Laboratory under DOE Contract No. DEFG02-96ER45439.


\begin{thebibliography}{10}

\bibitem{HOLZ94}
M.~H. H{\"o}lzle, U. Retter, and D.~M. Kolb, J. Electroanal.
  Chem., 371 (1994) 101.

\bibitem{ZHAN96}
J. Zhang, Y.-E. Sung, P.~A. Rikvold, and A. Wieckowski, J. Chem.
  Phys., 104 (1996) 5699.

\bibitem{SCHU76}
J.~W. Schultze and D. Dickertmann, Surf. Sci., 54 (1976) 489.

\bibitem{HACH91}
T. Hachiya, H. Honbo, and K. Itaya, J. Electroanal. Chem.,
  315 (1991) 275.

\bibitem{MAGN91}
O.~M. Magnussen {\it et~al.}, J. Vac. Sci. Tech. B, 9 (1991) 969.

\bibitem{TONE95}
M.~F. Toney {\it et~al.}, Phys. Rev. Lett., 75 (1995) 4472.

\bibitem{HUCK91}
D.~A. Huckaby and L. Blum, J. Electroanal. Chem., 315 (1991) 255.

\bibitem{RIKV88}
P.~A. Rikvold, J.~B. Collins, G.~D. Hansen, and J.~D. Gunton, Surf. Sci.,
203 (1988) 500.

\bibitem{BLUM90}
L. Blum, Adv. Chem. Phys., 78 (1990) 171.

\bibitem{GAMB93}
M. Gamboa-Aldeco {\it et~al.}, Surf. Sci., 297 (1993)  L135.

\bibitem{RIKV95}
P.~A. Rikvold {\it et~al.}, Surf. Sci., 335 (1995) 389.

\bibitem{RIKV91}
P.~A. Rikvold, Electrochim. Acta, 36 (1991) 1689.

\bibitem{RIKV96}
P.~A. Rikvold, J. Zhang, Y.~E. Sung, and A. Wieckowski, Electrochim. Acta, 
41 (1996) 2175.

\bibitem{BLUM96}
L. Blum, D.~A. Huckaby, and M. Legault, Electrochim. Acta,
41 (1996) 2207.

\bibitem{RIKV97a} 
P.~A. Rikvold, G. Brown, M.~A. Novotny, and
A. Wieckowski, Colloids and Surfaces A, 134 (1998) 3.

\bibitem{CHAP} 
G. Brown, P.~A. Rikvold, S.~J. Mitchell and M.~A. Novotny,
in {\em Interfacial Electrochemistry}, edited by A. Wieckowski, Marcel
Dekker (New York, in press).

\bibitem{BARD}
A.~J. Bard and L.~R. Faulkner, {\em Electrochemical Methods: Fundamentals and
  Applications}, Wiley (New York, 1980).

\bibitem{VATT96}
I. Vattulainen, J. Merikoski, T. Ala-Nissil{\"a}, and S.~C. Ying, Surf. Sci.,
366 (1996) L697.

\bibitem{NOVO95}
M.~A. Novotny, Computers in Physics, 9 (1995) 46.

\bibitem{RIKV97}
P.~A. Rikvold, A. Wieckowski, and R.~A. Ramos, Mater. Res. Soc. Symp. Proc.
  Ser., 451 (1997) 69.

\end{thebibliography}

\newpage
\pagestyle{empty}
~
\begin{figure}[tbp]
\vskip 3.1in
\includegraphics{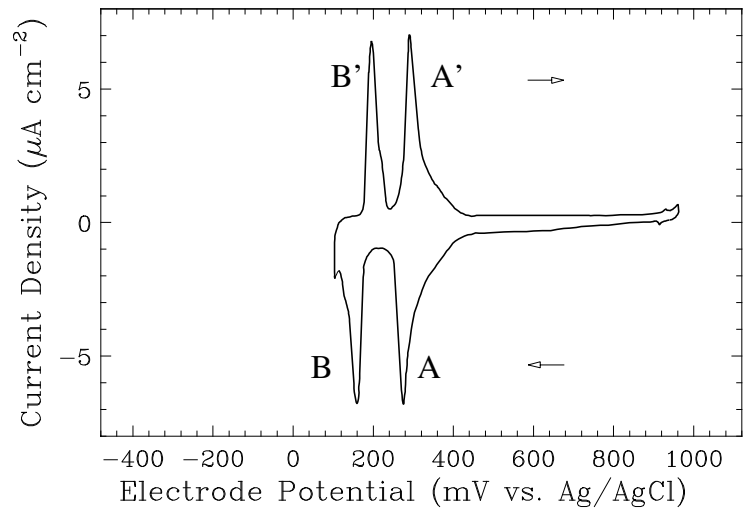}
\label{fig:CV}
\bigskip
\caption{Cyclic voltammogram for Cu UPD onto Au(111) surface,
after Ref.~\protect{\cite{ZHAN96}}. The two peaks correspond to two
first-order phase transitions in the adsorbed layer.  The arrows
indicate the direction of the potential scan.}
\end{figure}
\vfill

~
\begin{figure}[tbp]
\vskip 2.5in
\includegraphics{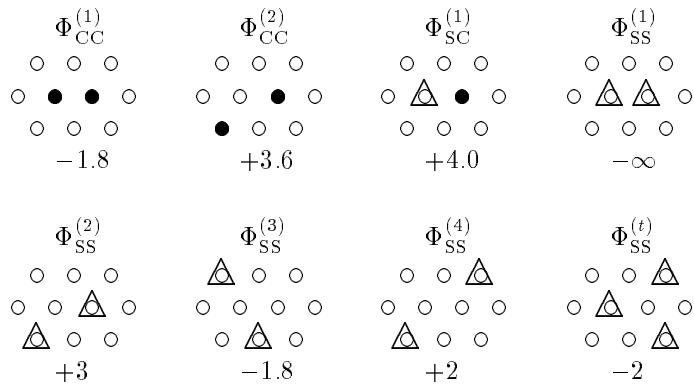}
\label{fig:PhiValues}
\bigskip
\caption{Diagrams illustrating the relative adsorbate positions
corresponding to the $n$-th neighbor interactions, $\Phi_{\rm
XY}^{(n)}$, in the lattice-gas model. Here Cu is represented by a
filled circle ({\large$\bullet$}), sulfate is represented by a
triangle ($\triangle$), and unoccupied sites are open circles
({\large$\circ$}). The interaction energy in  ${\rm kJ/mol}$ is given below each
diagram. Positive values indicate attractive interactions.
First-neighbor sulfates are subject to an infinitely strong repulsion,
{\em i.e.\/} configurations with first-neighbor sulfates are
forbidden. The lower right-most diagram illustrates the three-body
sulfate interaction. Note that some of the interactions are different
from those used in Ref.~\protect{\cite{ZHAN96}}.}
\end{figure}
\vfill

~
\begin{figure}[tbp]
\vskip 6in
\includegraphics{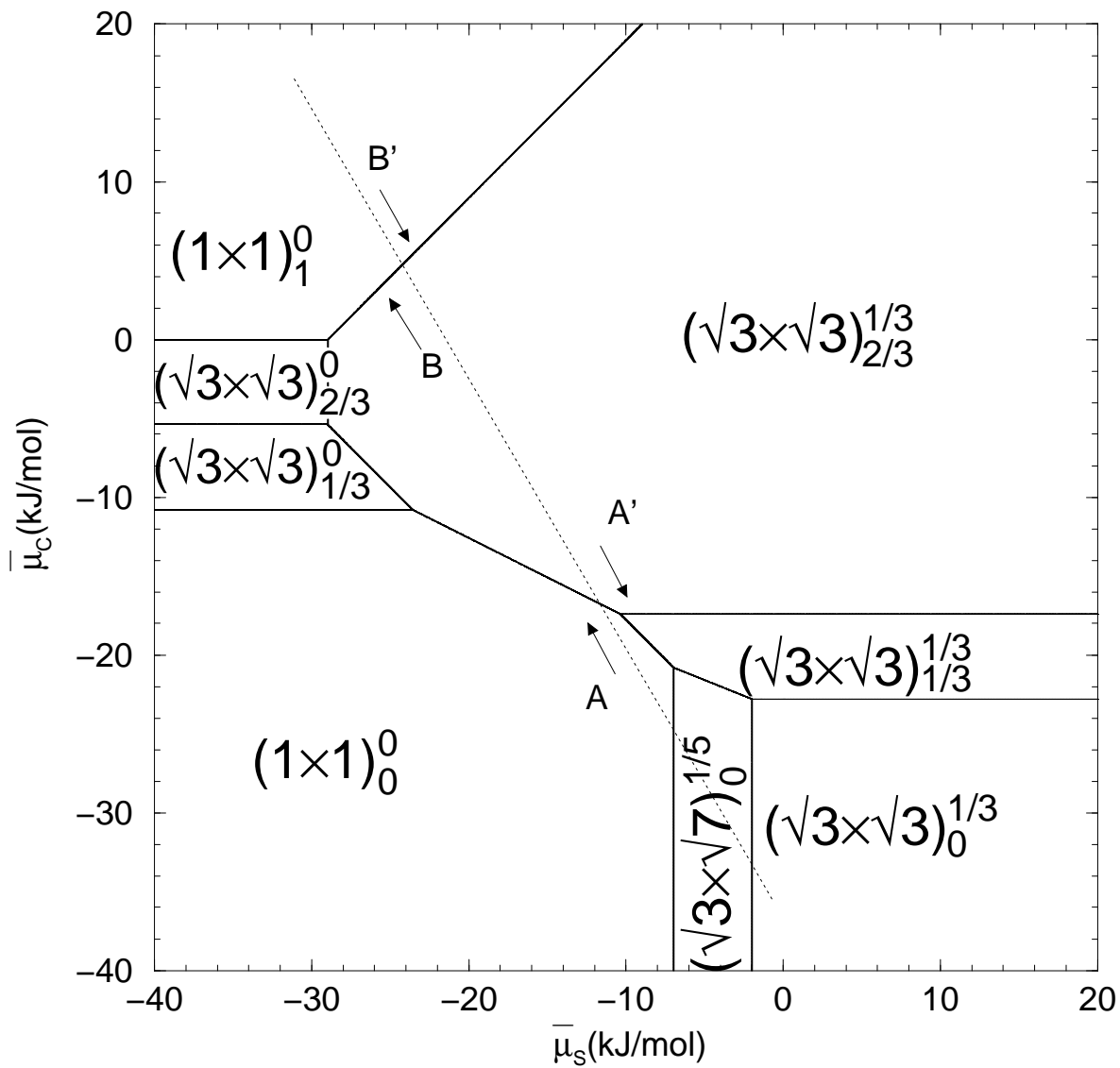}
\label{fig:DiagramIso}
\bigskip
\caption{The ground-state phase diagram (solid lines) for the
lattice-gas model with interactions as given in Fig.~2. The axes
represent the electrochemical potentials of the sulfate and copper,
$\bar{\mu}_{\rm S}$ and $\bar{\mu}_{\rm C}$, respectively. The
isotherm (dotted line) used for the simulated potential-step
experiments was chosen to match the transition potentials observed in
the experiments of HRK \protect{\cite{HOLZ94}}. The geometries of
the different phases are illustrated in Fig.~4.}
\end{figure}
\vfill

~
\begin{figure}[tbp]
\vskip 2.85in
\includegraphics{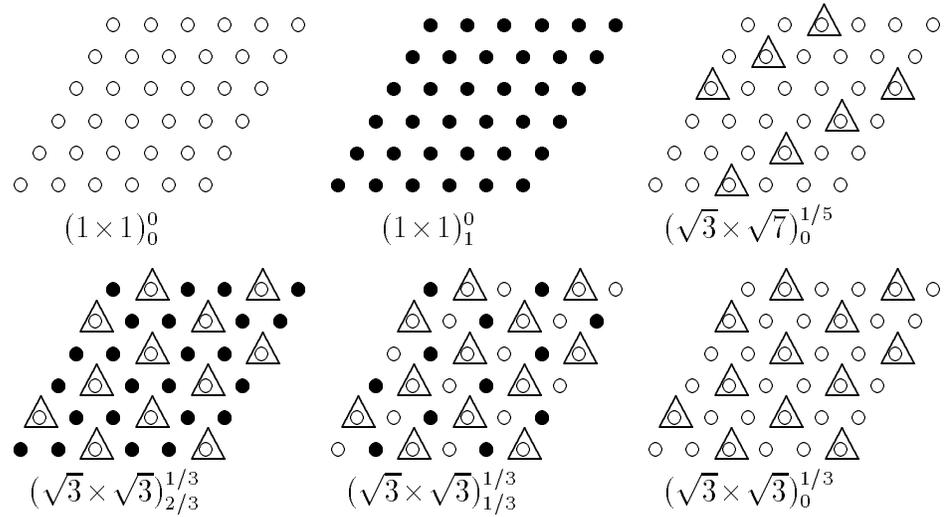}
\label{fig:PhaseExpl}
\bigskip
\caption{Important phases observed for the present lattice-gas model
of copper and sulfate adsorption on Au(111).  Cu is represented as a
filled circle ({\large$\bullet$}), sulfate as a triangle
($\triangle$), and empty sites as an unfilled circle
({\large$\circ$}). The label indicates the length of the unit vectors.
The subscript and superscript represent the fraction of sites occupied
by copper and sulfate, respectively. Two sulfate-free copper phases,
$(\sqrt{3}\times\sqrt{3})_{1/3}^0$ and
$(\sqrt{3}\times\sqrt{3})_{2/3}^0$, occur for this model at large
negative $\bar{\mu}_{\rm S}$, but are not shown. Their geometries are
easily deduced from the phases that are shown.}
\end{figure}
\vfill

~
\begin{figure}[tbp]
\vskip 3in
\includegraphics{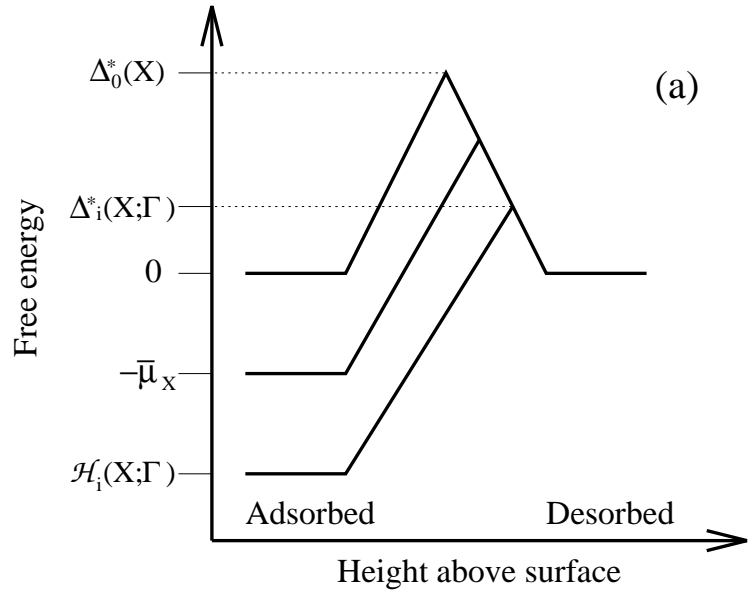}
\bigskip
\vskip 3in
\includegraphics{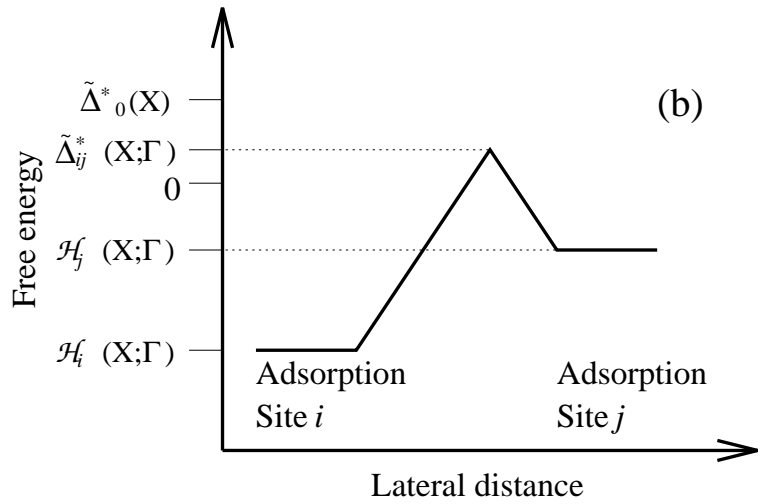}
\label{fig:Barriers}
\bigskip
\caption{Butler-Volmer type free-energy barrier scheme. See text for
details.}
\end{figure}
\vfill

~
\begin{figure}[tbp]
\vskip 3in
\includegraphics{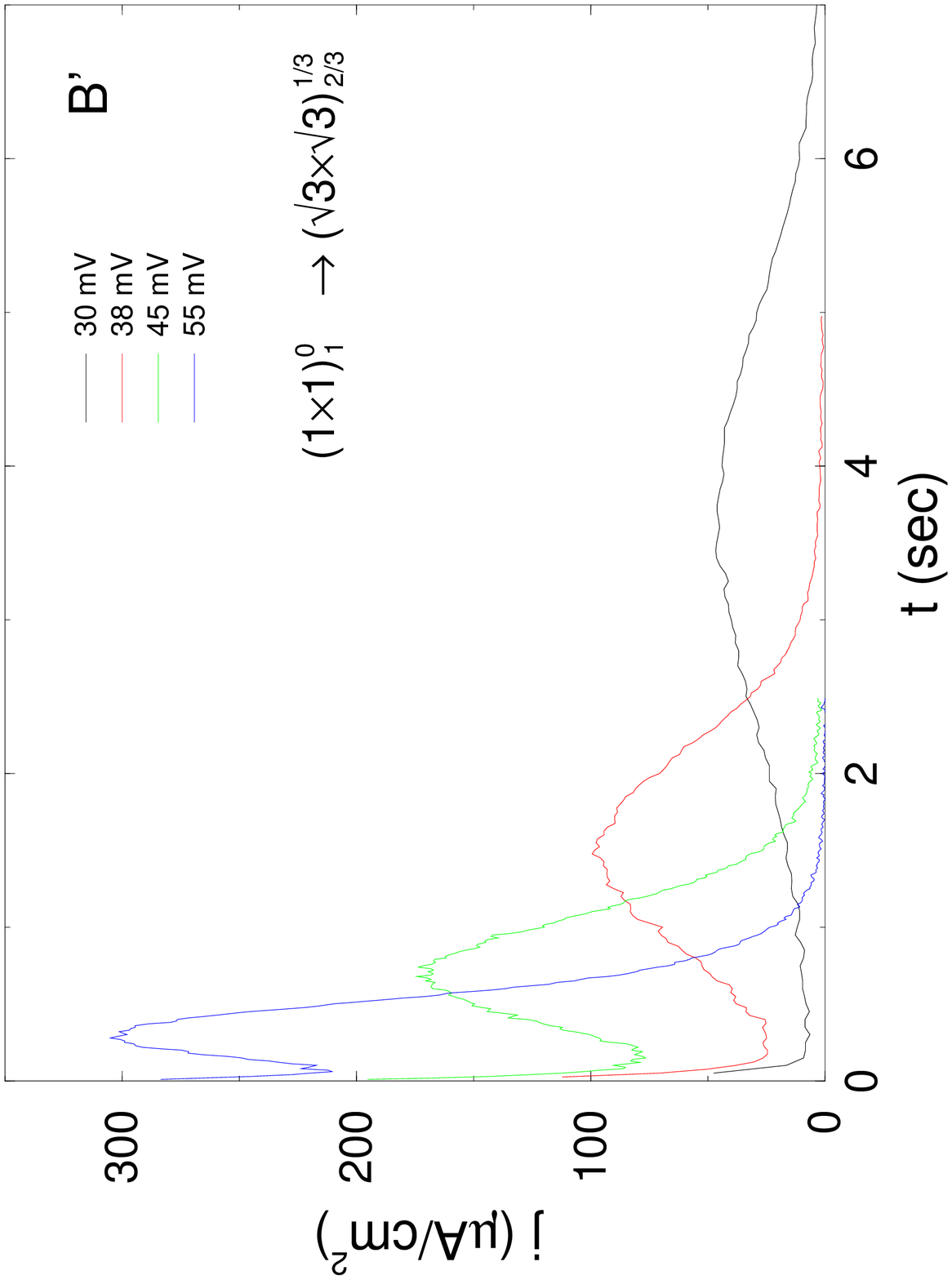}
\label{fig:CurrentBp}
\bigskip
\vskip 3in
\includegraphics{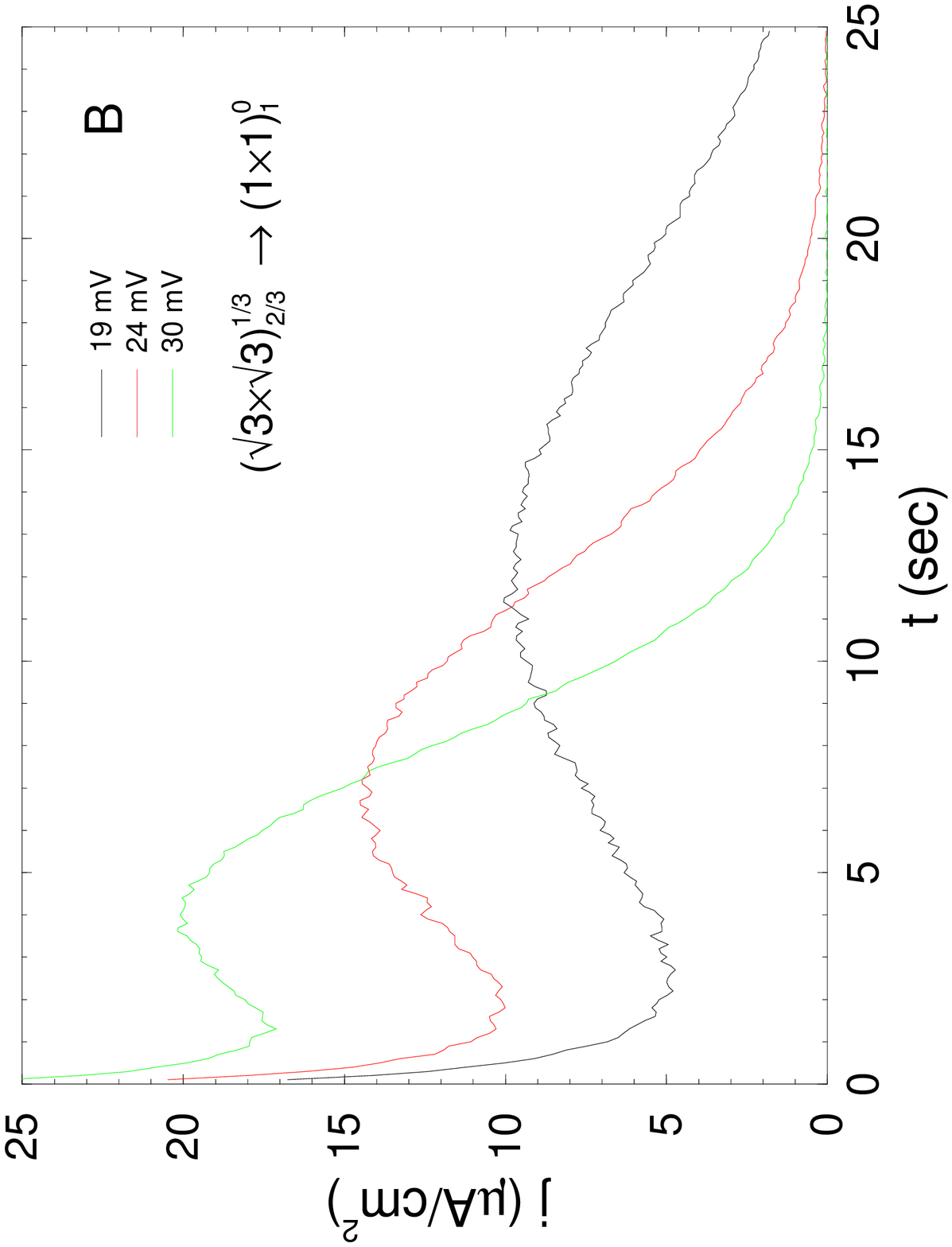}
\label{fig:CurrentB}
\bigskip
\caption{Simulated current transients after potential steps across the
transition between the $(1\times 1)^0_1$ and
$(\sqrt{3}\times\sqrt{3})^{1/3}_{2/3}$ phases. Positive (top) and
negative (bottom) potential steps, ${\rm B}'$ and ${\rm B}$,
respectively, of HRK \cite{HOLZ94} are shown.  The shape of the
current profile corresponds to the decay of a metastable adsorbed
layer. The trend for the broad maxima to shift to earlier times and
larger amplitudes as the potential step is increased agrees with the
experiments. The legend indicates the size of the potential step past
the transition, $|E_{\rm final}-E_{\rm transition}|$. This figure
should be compared with Fig.~3 of Ref.~\protect{\cite{HOLZ94}}.}
\end{figure}
\vfill

~
\begin{figure}[tbp]
\vskip 3in
\includegraphics{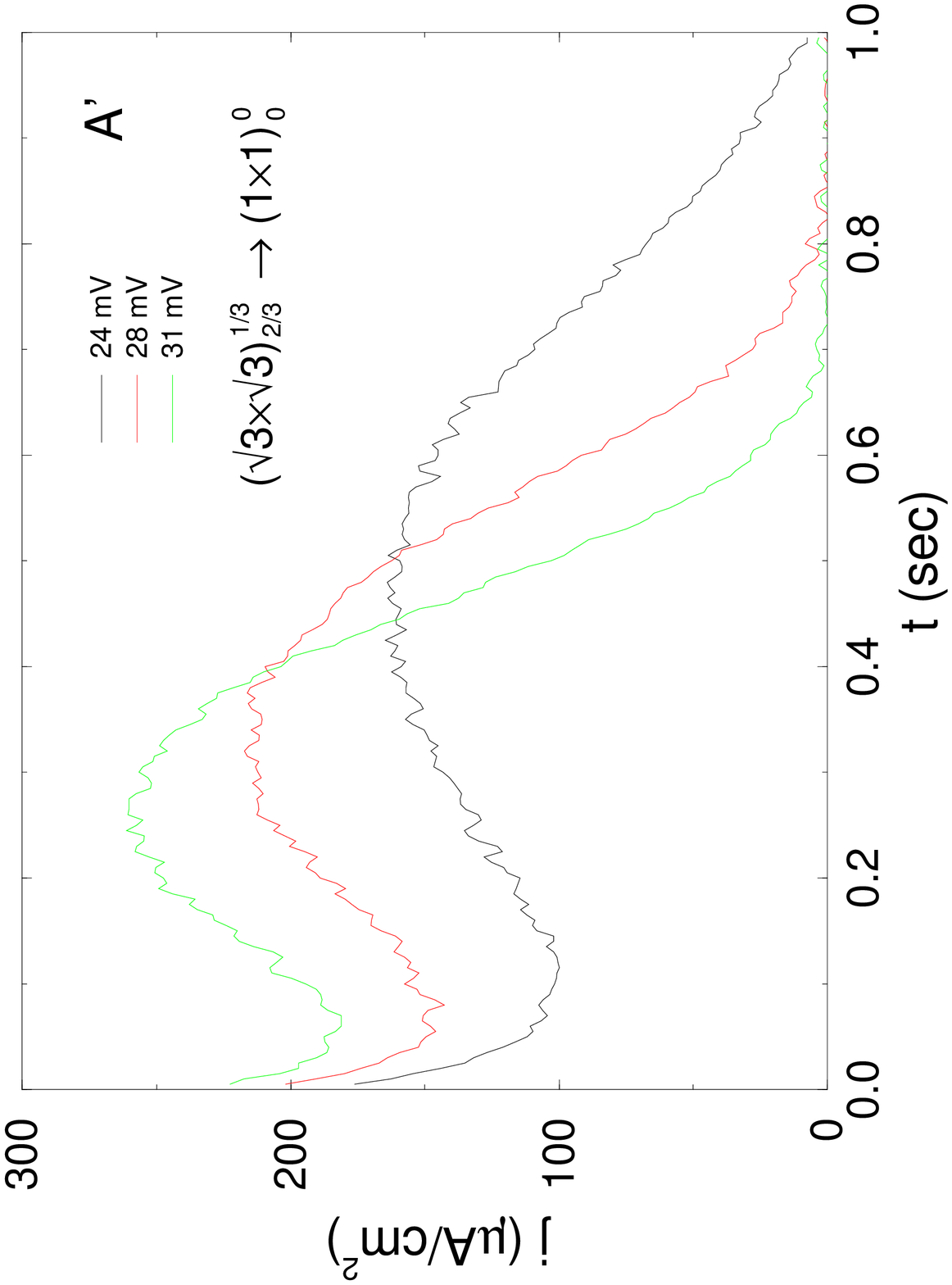}
\bigskip
\vskip 3in
\includegraphics{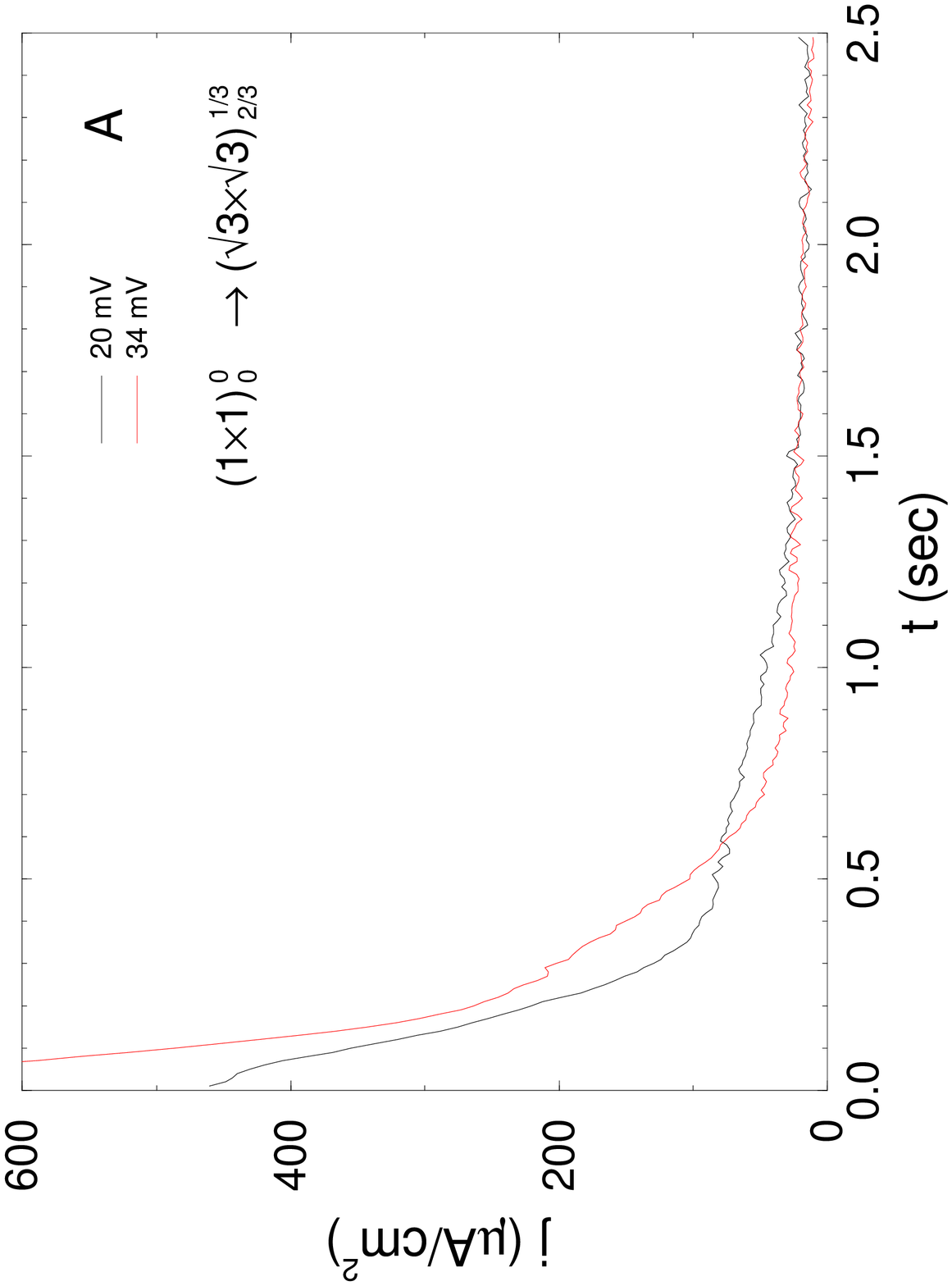}
\label{fig:CurrentA}
\bigskip
\caption{Simulated current transients after potential steps across the
disorder-mixed transition for positive (top) and negative (bottom)
potential steps, ${\rm A}'$ and ${\rm A}$, respectively, of HRK
\cite{HOLZ94}.  The strong asymmetry in current profiles for the two
step directions is seen experimentally. The broad maximum is
characteristic of the decay of a metastable adsorbed layer, the
monotonically decreasing initial current reflects the unstable nature
of the low-coverage surface after a negative potential step. The
legend indicates the size of the potential step past the transition,
$|E_{\rm final}-E_{\rm transition}|$. This figure should be compared
with Fig.~3 of Ref.~\protect{\cite{HOLZ94}}.}
\end{figure}
\vfill

~
\begin{figure}[tbp]
\vskip 7in
\includegraphics{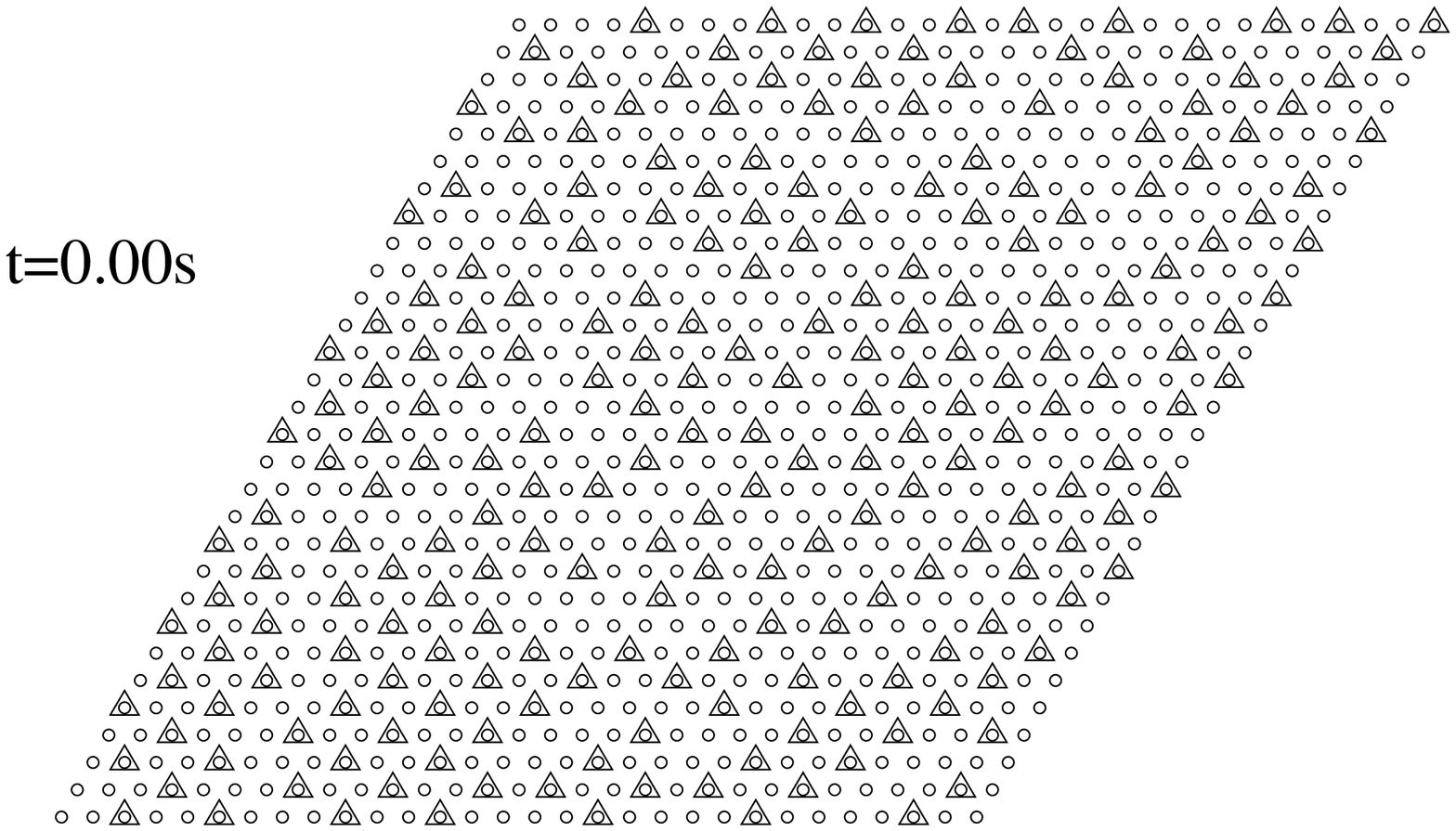}
\includegraphics{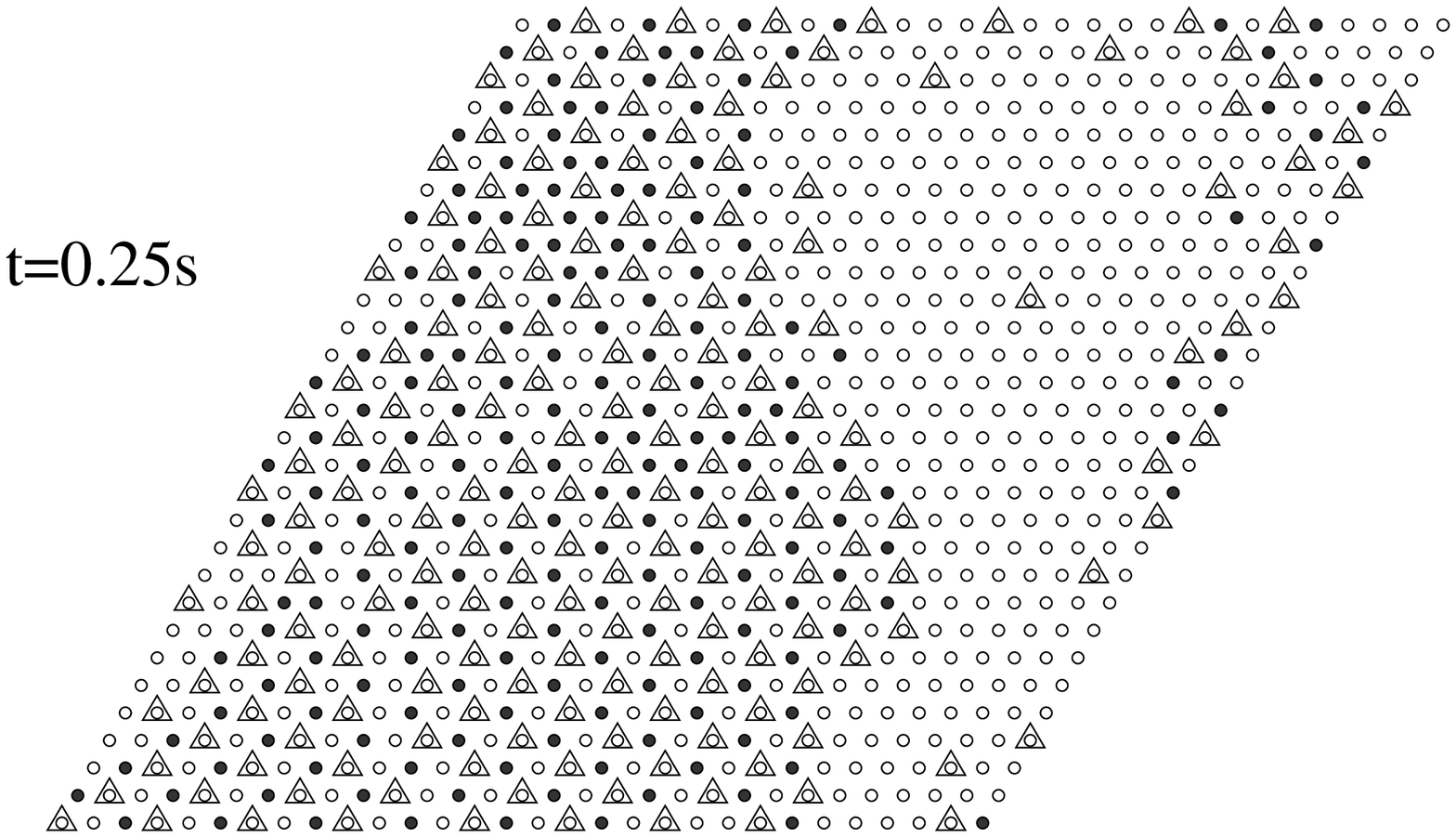}
\includegraphics{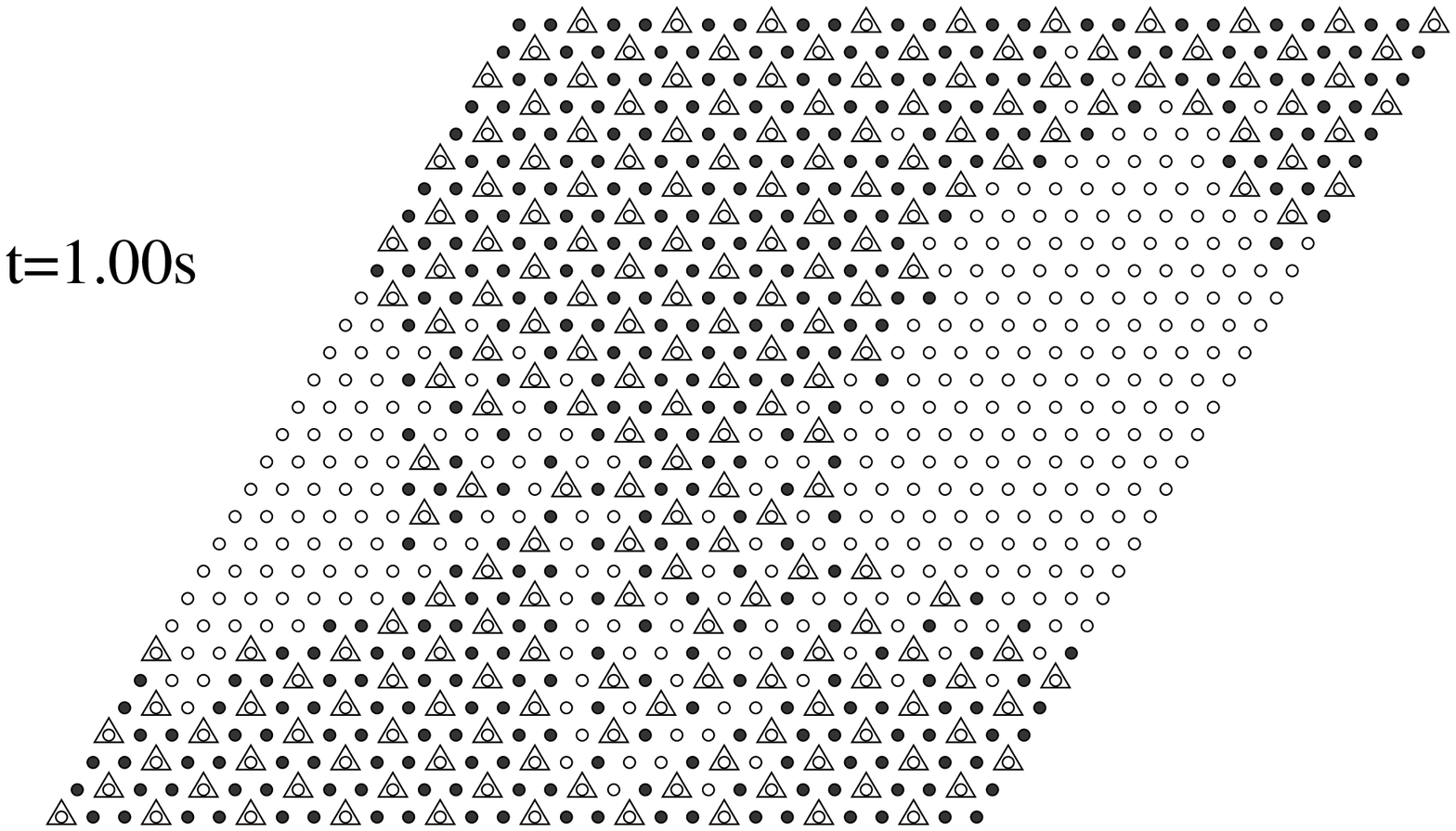}
\label{fig:ConfigA}
\bigskip
\caption{A series of snapshots after a negative-going potential step
to 20mV below the transition between the low coverage and mixed
layers. After the step, a fraction of the sulfate desorbs, but the
remaining sulfate combines with the newly adsorbed copper to form a
loose domain. With time this domain fills in and grows. The current
transient is monotonically decreasing, as shown in Fig.~7,
${\rm A}$. Movies of the dynamics after the quench are available in MPEG
format at \hbox{\tt http://www.scri.fsu.edu/\~{ }rikvold} }
\end{figure}
\vfill

\end{document}